\documentclass[12pt]{iopart}

\expandafter\let\csname equation*\endcsname\relax

\expandafter\let\csname endequation*\endcsname\relax

\usepackage{amsmath}
\usepackage{graphicx}
\usepackage{dcolumn}
\usepackage{braket}

\def\bcen{\begin{center}}
\def\ecen{\end{center}}
\usepackage{epstopdf} 
\begin{document}

\title{A proposal for the implementation of quantum gates in an optomechanical system \textit{via} phonon blockade}

\author{Nilakantha Meher}
\address{Department of Physics, Indian Institute of Technology Kanpur, Kanpur 208016, India}
\ead{nilakantha.meher6@gmail.com}
\vspace{10pt}

\begin{abstract}

We propose a scheme to implement quantum controlled NOT gate and quantum phase gate in an optomechanical system based on phonon blockade.  For appropriate choices of system parameters, fidelities of both the quantum gate operations are very close to unity in the absence of dissipation. Using recently achieved experimental values in a mechanical resonator coupled to a microwave cavity, we show that the quantum gate can be realized experimentally with very high fidelity.
\end{abstract}

%
\vspace{2pc}
%
\submitto{\JPB}
%
\maketitle
%
%

\section{Introduction}
Realizing quantum computer in a physical system is of interest in quantum information processing and computation \cite{ DiVincenzo4, DiVincenzo3, Ladd, Barz, Cai}. The idea of quantum computation was first proposed by Benioff \cite{Beni} and further developed by Feynman \cite{Feyn}. Deutsch and Josza showed that a quantum computer can solve a problem much faster than a classical computer and they introduced logic gate operations \cite{Deut}. The number of operations required to implement various algorithms is reduced exponentially in a quantum computer compared with a classical computer. Quantum logic gates are implemented for such quantum computation and information processing \cite{Shor, Fredkin, DiVincenzo2, Bare}. Moreover, a sequence of quantum gates is wired together to form a quantum circuit for solving a desired problem. A set of quantum gates is said to be universal if any quantum circuit can be constructed using this set of gates \cite{Fredkin, DiVincenzo2,Bare, Sleator, Lloy, David, Buzek2,Shi, Neil}. For instance, a set of unitary single qubit gates with controlled NOT (CNOT) gate, \textit{i.e,} $\{$CNOT, Hadamard, phase and $\pi/8 (T)\}$, collectively form a universal set. Hence, realizing these gates in a physical system is a basic requirement for quantum computation \cite{DiVincenzo4,DiVincenzo3}. So far, quantum gates have been realized in various physical systems such as nuclear magnetic resonance \cite{Gers, Jones}, trapped ions \cite{Cirac, Monroe}, photons in polarization degree of freedom \cite{Knil, Brie}, cavity quantum electrodynamics \cite{Sleator, Domokos, Turchette, Rauschenbeutel,Giovannetti, Li, Su, Wei, Wei2, Wang}, quantum dots \cite{Loss, HuCY, HuCY2}, superconducting resonators \cite{Hua, Hua2}, diamond nitrogen-vacancy centre \cite{Charnock,Kennedy, Su, Wei4}.  \\

An optomechanical system consists of a quantized electromagnetic field in a cavity interacting with the mechanical degree of freedom of an oscillator \textit{via} radiation pressure \cite{Aspe}. This radiation pressure results from the momentum carried by photon. Various theoretical and experimental efforts have been made to realize single photon source \cite{Rabl, Nunnenkamp}, quantum state transfer \cite{Jia,Neto,Zhang2, Xu}, entanglement generation \cite{Ren, Vitali}, bistability \cite{Veng}, mechanical squeezing \cite{Agar}, ground state cooling of mechanical motion \cite{Chan, Teuf}, optomechanically induced transparency \cite{Huan}, etc. in optomechanical systems. Recently, Simon and Hartmann introduced an approach to store the information in the mechanical degree of freedom and proposed a scheme to realize an ISWAP gate \cite{Rips}. A Controlled phase-flip gate is perceived in the polarization degree of freedom of cavity field \cite{Wen} and in two-mode field in an optomechanical system \cite{Muha}.\\

With the development of  fabrication technology, it is possible to fabricate cavities and mechanical resonators with high quality factor (Q-factor) \cite{Sun, Thompson, Chang2}.  This provides a suitable physical system to demonstrate many quantum optical phenomena.  Moreover, recent experimental techniques allow for controlling the optomechanical coupling at the single photon level and the resonator can be coupled to electromagnetic fields in a very broad frequency range \cite{Rips}. Hence, it is of interest to realize quantum gates in an optomechanical system where the cavity state and mechanical state are easily controllable and measurable. In this article, we propose a scheme to realize CNOT gate and quantum phase gate (QPG) in an optomechanical system. Our model consists of a mechanical resonator which is of Kerr type.  
For appropriate choices of nonlinear strength, resonance frequencies and cavity-resonator coupling strength, fidelity of the gate operation is very close to unity. We also investigate the gate fidelity in the presence of photon loss and mechanical damping. \\

This article is organized as follows: In Sec. \ref{PhysicalModel}, we introduce our model and discuss the phonon blockade mechanism in the model. We discuss the theory of the CNOT gate and QPG in Sec. \ref{OPtasGate}. Results are discussed using experimentally accessible parameters in Sec. \ref{Effectofdissipation}. In Sec. \ref{HigherStates}, the effect of higher excited states of mechanical resonator on gate fidelity is considered. We summarize our results in Sec. \ref{Summary}

\section{Physical model}\label{PhysicalModel}
We consider a cavity interacting with a mechanical resonator with coupling strength $g$. The Hamiltonian for the system is \cite{Liu}
\begin{align}\label{H}
H=\omega_ca^\dagger a+\omega_m b^\dagger b-\chi (b^{\dagger }b)^2+ga^\dagger a(b+b^\dagger).
\end{align} 
Here $a^\dagger (a)$ is the creation (annihilation) operator for the cavity and $b^\dagger (b)$ is the creation (annihilation) operator for the mechanical resonator. The resonance frequencies of the cavity and mechanical resonator are $\omega_c$ and $\omega_m$ respectively. The mechanical resonator is considered to be of Kerr type with strength $\chi$ \cite{Rips2, Stan, Liu}.
A key feature of the Hamiltonian given in Eqn. \ref{H} is that it conserves the  total number of photons, \textit{i.e.} $[a^\dagger a,H]=0$. Thus, $H$ is a block diagonal matrix for each photon number subspace.\\

\begin{figure}
\centering
\includegraphics[scale=0.4]{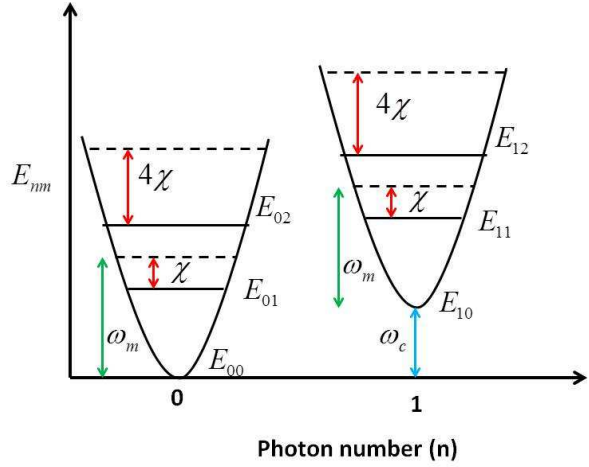}
\caption{Energy levels $E_{nm}$ for $n=0,1$ and $m=0,1,2$. Here $n$ represents the number of photon in the cavity and $m$ is the number of phonon in the resonator.}
\label{Energylevel}
\end{figure}

For $g=0$, the state $\ket{n,m}$ are the eigenstates of the Hamiltonian $H$ and the corresponding eigenvalues are $E_{nm}=n\omega_c+m\omega_m-\chi m^2$. Here $n$ represents the number of photons in the cavity and $m$ is the number of phonons in the resonator. The energy levels $E_{nm}$ for various values of $n$ and $m$ are shown in Fig. \ref{Energylevel}. If $\chi=0$, the energy levels are harmonic. 
Anharmonicity arises due to the presence of Kerr nonlinearity, \textit{i.e.,} $\chi \neq 0$. This is the basic requirement for phonon blockade to occur.
The energy difference between two nearest energy levels for a given $n$ is
\begin{align*}
\Delta E_{n,k}=E_{n,k+1}-E_{n,k}=(\omega_m-\chi)-2\chi k.
\end{align*}
Interestingly, for $\omega_m=\chi$, the ground state and the first excited state of the mechanical resonator are degenerate.  We set $\omega_m=\chi$ in subsequent discussion. Also, we restrict the number of photon in the cavity to be either 0 or 1.\\ 

\begin{figure}[h]
\centering
\includegraphics[scale=0.3]{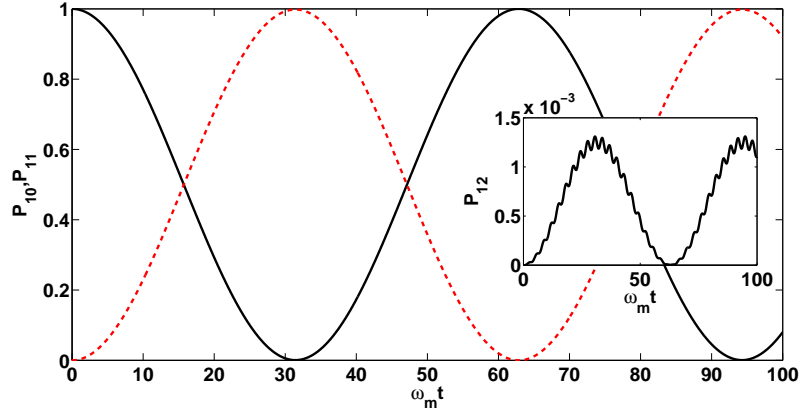}
\caption{Probabilities of detecting the state $\ket{10}$(solid line) and $\ket{11}$(dashed line) as a function of $\omega_mt$. The inset figure shows the probability of detecting two phonons ($P_{12}$) in the mechanical resonator. We set $g/\omega_m=0.05$ and $\chi/\omega_m=1$.  }
\label{phononblockade}
\end{figure}

For $n=0$, \textit{i.e.,} if the cavity is in vacuum state, the state $\ket{0,m}$ are the eigenstates of $H$. However, for $n=1$ and $g\neq 0$, the approximate eigenstates of $H$ are (for $\chi>>g$)
\begin{subequations}\label{eigenstate}
\begin{eqnarray}
\ket{X_+}&\approx\frac{1}{\sqrt{2}}(\ket{10}+\ket{11})+\frac{g}{2\chi}\ket{12},\\
\ket{X_-}&\approx\frac{1}{\sqrt{2}}(\ket{10}-\ket{11})-\frac{g}{2\chi}\ket{12},\\
\ket{X_{12}}&\approx \ket{12}-\frac{g}{\sqrt{2}\chi}\ket{11},
\end{eqnarray}
\end{subequations} 
with overall normalization factors. The corresponding eigenvalues are $\omega_c+ g$, $\omega_c-g$ and $-2\chi$ respectively. Here $\ket{nm}$ represents $n$ photons in the cavity and $m$ phonons in the mechanical resonator. The contributions from higher excited states, such as $\ket{1m}$ where $m>1$, are negligible. Hence the evolution of the state $\ket{10}$ under the Hamiltonian given in Eqn. \ref{H} is
\begin{align}
e^{-iHt}\ket{10}\approx e^{-i\omega_c t}[\cos gt \ket{10}-i\sin gt\ket{11}]+\mathcal{O}\left(g/\sqrt{2}\chi\right).
\end{align} 
The probabilities of detecting the states $\ket{10}$ and $\ket{11}$ during time evolution are shown in Fig. \ref{phononblockade}. It is to be noted that the mechanical resonator oscillates between its ground state and first excited state. Hence, the mechanical resonator acts as a two-level system. The probability of detecting two phonons ($P_{12})$ in the mechanical resonator is of the order of $\sim g^2/2\chi^2$ which is negligible if $g<<\chi$ (refer to inset of Fig. \ref{phononblockade}). Essentially, anharmonicity in the energy levels inhibits the transition of mechanical resonator to higher excited states if $\omega_m=\chi>>g$, which is referred to as a \textit{phonon blockade} \cite{Liu}.  

\section{Optomechanical cavity as quantum gates}\label{OPtasGate}
In this section, we discuss the CNOT gate and QPG in the optomechanical cavity. The state of the cavity is considered to be the control bit whereas the state of the mechanical resonator is the target bit.
\subsection{Controlled NOT gate}
The CNOT gate operates on a two qubit state.  Mathematically, the operation for the CNOT gate is
$\ket{a}\ket{b}\rightarrow \ket{a}\ket{a\oplus b}$, 
with $a,b \in \{0,1\}$ and $\oplus$ denotes addition modulo 2. Here $\ket{a}$ is the control bit and $\ket{b}$ is the target bit. In the CNOT operation, the state of the target bit remains unchanged if the state of the control bit is $\ket{0}$. Flipping occurs between $\ket{0}$ and $\ket{1}$ of the target bit if the state of the control bit is $\ket{1}$.\\
 
Now, we consider the evolution of states $\ket{00},\ket{01}, \ket{10}$ and $\ket{11}$ under the Hamiltonian $H$ given in Eqn. \ref{H}. The time evolved states, in the limit $\chi>>g$, are
\begin{align*}
e^{-iHt}\ket{00} &= \ket{00},\\
e^{-iHt}\ket{01} &= \ket{01},\\
e^{-iHt}\ket{10} &= e^{-i\omega_c t}[\cos gt \ket{10}-i\sin gt\ket{11}],\\
e^{-iHt}\ket{11} &= e^{-i\omega_c t}[\cos gt \ket{11}-i\sin gt\ket{10}].
\end{align*}
These unitary evolutions become CNOT operation if $\omega_c/g=4k-1$ and $t=\pi/2g$, where $k$ is a non-zero integer. The time $t=\pi/2g$ is the gate operation time which we denote as $T_{G}$. 
\subsection{Quantum phase gate}
This is the simplest two qubit conditional gate, where the phase of the target bit changes conditionally. The QPG operation is
$\ket{a,b}\rightarrow \exp(-i\phi \delta_{a,1} \delta_{b,1})\ket{a,b},$
where $\delta_{a,1}$ and $ \delta_{b,1}$ are Kronecker delta symbols and $a,b \in \{0,1\}$.\\
 
We consider the evolution of the states $\ket{0}\ket{-}, \ket{0}\ket{+}, \ket{1}\ket{-}, $ and $\ket{1}\ket{+}$. These states evolve under the Hamiltonian $H$ as
\begin{align*}
e^{-iHt}\ket{0}\ket{-}&=\ket{0}\ket{-},\\
e^{-iHt}\ket{0}\ket{+}&=\ket{0}\ket{+},\\
e^{-iHt}\ket{1}\ket{-}&=e^{-i(\omega_c-g)t}\ket{1}\ket{-},\\
e^{-iHt}\ket{1}\ket{+}&=e^{-i(\omega_c+g)t}\ket{1}\ket{+},
\end{align*}
where $\ket{\pm}=\frac{1}{\sqrt{2}}(\ket{0}\pm \ket{1})$ is the superposition of the ground state and single phonon state of the mechanical resonator. At $t=\frac{2p\pi}{\omega_c-g}$, the unitary evolution becomes a QPG operation if
\begin{align*}
\frac{\omega_c}{g}=\frac{2(p+q)\pi+\phi}{2(q-p)\pi+\phi},
\end{align*}
where $p$ and $q$ are two integers. For this case, $T_G=\frac{2p\pi}{\omega_c-g}$ is the gate operation time.
\section{Results and discussions}\label{Effectofdissipation}
As seen in the previous section, transition between the mechanical resonator states requires the presence of a photon in the cavity. Hence, to turned on the quantum gates, it is required to load a photon into the cavity \cite{Su, Liu2}. Once the photon is inside the cavity, it interacts with the mechanical resonator.  Then the optomechanical system works like a quantum gate for appropriate choices of system parameters such as resonance frequencies, nonlinear strength and coupling strength. Finally, it is required to turn off the gate at a specific time, so called gate operation time, to obtained the desired results. To turn off the gate, one needs to remove the photon from the cavity \cite{Su}. However, in experimental situation the cavity is not ideal and there are finite probability of leakage of the photon from the cavity before the gate operation is complete. In such case, the fidelity of the quantum gate will be less. Hence, it is necessary to investigate the effect of dissipation on quantum gate fidelity.     \\

In this section, we discuss CNOT gate and QPG operations using the realistic values of system parameters. Dissipation is being included. Here, dissipation leads to photon loss from the cavity and damping of the mechanical resonator. The effect of dissipation on quantum gate fidelity is studied by analyzing the master equation. The master equation, which includes dissipation of both the cavity and mechanical resonator is \cite{Carmichael}
\begin{align}\label{Master}
\frac{\partial \rho}{\partial t}=\frac{-i}{\hbar}[H,\rho]+\frac{\gamma_c}{2}\mathcal{L}(a)\rho+\frac{\gamma_m}{2}\mathcal{L}(b)\rho,
\end{align}
where
\begin{align*}
\mathcal{L}(o)\rho=(2o\rho o^\dagger-o^\dagger o \rho-\rho o^\dagger o),
\end{align*}
is the Lindblad superoperator \cite{Lindblad}.
Here $\gamma_c$ and $\gamma_m$ are the decay rates of the cavity  and the mechanical resonator respectively. It is known that the Lindblad master equation is valid if the system-reservoir couplings are weaker than the intra-system coupling, and also these couplings are weaker than the resonance frequencies \cite{Gonzalez, Rivas, Qing, Wang2}. For strong couplings, one may use a global master equation to describe the evolution of the system \cite{Chou}. \\

 The effectiveness of quantum gates is characterized by fidelity between initial state and target state. The fidelity between two states $\rho$ and $\sigma$ is defined as \cite{Jozsa}
\begin{align}\label{FidelityMaster}
F=\left[\text{Tr}\sqrt{\sqrt{\sigma}\rho\sqrt{\sigma}}\right]^2.
\end{align}

We discuss the CNOT gate first and then QPG gate. Let the initial state of the system be
\begin{align}\label{inputCNOT}
\ket{\psi_{in}}=c_1\ket{00}+c_2\ket{01}+c_3\ket{10}+c_4\ket{11},
\end{align}
where $\sum_{i=1}^4|c_i|^2=1$. This is the most general two qubit state and any two qubit state can be obtained by suitably choosing the superposition coefficients. If the CNOT operation works for this state, then it will work for any two-qubit state. The action of the CNOT gate on $\ket{\psi_{in}}$ will generate the state
$\ket{\Psi}=c_1\ket{00}+c_2\ket{01}+c_3\ket{11}+c_4\ket{10}$. Using the definition given in Eqn. \ref{FidelityMaster}, the fidelity for the CNOT gate is 
\begin{align}\label{CNOTFidelity}
F_{CNOT}(T_G)=\left[\text{Tr}\sqrt{\sqrt{\sigma}\rho(T_G)\sqrt{\sigma}}\right]^2,
\end{align}
where $\rho(T_G)$ is the state of the system at time $T_G=\pi/2g$ evolved from the state given in Eqn. \ref{inputCNOT}. The evolution of the system is governed by Eqn. \ref{Master}. Here $\sigma=\ket{\Psi}\bra{\Psi}$. $F_{CNOT}=1$ corresponds to perfect CNOT gate. \\
\begin{figure}[h]
\centering
\includegraphics[scale=0.3]{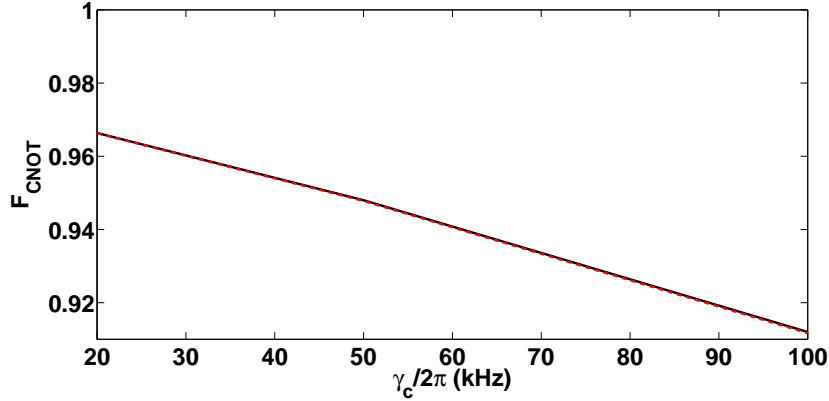}
\caption{ Average fidelity $F_{CNOT}$ as a function of $\gamma_c/2\pi$. The average is taken over the ensemble of initial  states given in Eqn. \ref{inputCNOT}, where $c_i$'s are randomly chosen. The continuous line is the average fidelity calculated using Eqn. \ref{outputCNOT} and dashed line is the average fidelity calculated by solving master equation numerically. We use $\omega_m/2\pi=15.9 \times 10^6$ Hz, $g/2\pi=242\times 10^3$Hz, $\chi=\omega_m$, $\omega_c/2\pi=2.4197\times 10^9$ Hz and $\gamma_m/2\pi=150$ Hz.}
\label{averagefidelity}
\end{figure}

Using the currently available experimental values in a mechanical resonantor coupled to a microwave cavity \cite{Lecocq}, we calculate $F_{CNOT}$ using Eqn. \ref{CNOTFidelity}.  We set $\omega_m/2\pi=15.9$ MHz, $g/2\pi=242$ kHz, $\gamma_m=150$ Hz  and $\omega_c=2.4197$ GHz. This gives the CNOT gate operation time $T_G \sim 1 \mu$s. We plot $F_{CNOT}$ as a function of cavity decay rate $\gamma_c/2\pi$ in Fig. \ref{averagefidelity}. The dashed line represents the average fidelity calculated from Eqn. \ref{CNOTFidelity}. The average is taken over the ensemble of initial  states given in Eqn. \ref{inputCNOT}, where $c_i$'s are randomly chosen.\\

In the presence of dissipation, the state of the system does not remain pure and hence, the pure state evolution is not possible. However, an approximate evolved state can be calculated by including phenomenological dissipation in the system. The evolved state, for $\chi>>g$, is
\begin{align}\label{outputCNOT}
\ket{\psi(t)}\approx \tilde{c}_1\ket{00}+\tilde{c}_2\ket{01}+\tilde{c}_3 \ket{10}+\tilde{c}_4\ket{11},
\end{align}
where $\tilde{c}_2=c_2 e^{-\gamma t/2},
\tilde{c}_3=e^{-i\omega_c t} e^{-\gamma t/2}\left(c_3 \Omega_+-ic_4 \Omega_-\right),
\tilde{c}_4=e^{-i\omega_c t} e^{-\gamma t/2}\left(c_4 \Omega_+-ic_3 \Omega_-\right),
\tilde{c}_1=\sqrt{1-|c_2|^2-|c_3|^2-|c_4|^2}e^{i \arg(c_1)}.$
Here $\Omega_+=\sqrt{\frac{1}{2}\left(1+e^{\frac{-3\gamma t}{4}}\cos 2gt\right)}$ and $\Omega_-=\sqrt{\frac{1}{2}\left(1-e^{\frac{-3\gamma t}{4}}\cos 2gt\right)}$. Then the fidelity becomes
\begin{align}\label{Fidelity}
F_{CNOT}=|\langle \Psi |\psi(T)\rangle|^2= |c_1^*\tilde{c}_1+c_2^*\tilde{c}_2+c_3^*\tilde{c}_4+c_4^*\tilde{c}_3|^2.
\end{align}
In Fig. \ref{averagefidelity}, the continuous line represents the average fidelity calculated using Eqn. \ref{Fidelity}, which is in exact agreement with the average fidelity calculated using the master equation (dashed line).\\

For the QPG gate, consider the input state
\begin{align}\label{inputQPG}
\ket{\psi_{in}}=c_1\ket{0}\ket{-}+c_2\ket{0}\ket{+}+c_3\ket{1}\ket{-}+c_4\ket{1}\ket{+},
\end{align}
with $\sum_{i=1}^4|c_i|^2=1$. Then the target state is
$\ket{\psi}=c_1\ket{0}\ket{-}+c_2\ket{0}\ket{+}+c_3\ket{1}\ket{-}+e^{-i\phi}c_4\ket{1}\ket{+}.$
The fidelity for the QPG gate is
\begin{align}\label{FidelityQPG}
F_{QPG}(T_G)=\left[\text{Tr}\sqrt{\sqrt{\sigma}\rho(T_G)\sqrt{\sigma}}\right]^2.
\end{align}
where $\rho(T_G)$ is the state of the system at time $T_G=\frac{2p\pi}{\omega_c-g}$ evolved from the state given in Eqn. \ref{inputQPG}. This evolution is calculated using Eqn. \ref{Master}. Here $\sigma=\ket{\Psi}\bra{\Psi}$.
 The approximate evolved state in the presence of dissipation is
\begin{align}
\ket{\psi(t)}&=\frac{d_1-d_2}{\sqrt{2}}\ket{0}\ket{-}+\frac{d_1+d_2}{\sqrt{2}}\ket{0}\ket{+}+\frac{d_3-d_4}{\sqrt{2}}\ket{1}\ket{-}+\frac{d_3+d_4}{\sqrt{2}}\ket{1}\ket{+},
\end{align}
where
$d_2= e^{-\gamma t/2}\frac{c_2-c_1}{\sqrt{2}},
d_3=e^{-i\omega_ct}e^{-\gamma t/2}\left[\Omega_+\frac{c_3+c_4}{\sqrt{2}}-i\Omega_-\frac{c_3-c_4}{\sqrt{2}}\right]$,\\
$d_4=e^{-i\omega_ct}e^{-\gamma t/2}\left[\Omega_+\frac{c_3-c_4}{\sqrt{2}}-i\Omega_-\frac{c_3+c_4}{\sqrt{2}}\right]$, $
d_1=\sqrt{1-|d_2|^2-|d_3|^2-|d_4|^2}e^{i \arg(\frac{c_1+c_2}{\sqrt{2}})}.$\\

\begin{figure}[h!]
\centering
\includegraphics[scale=0.3]{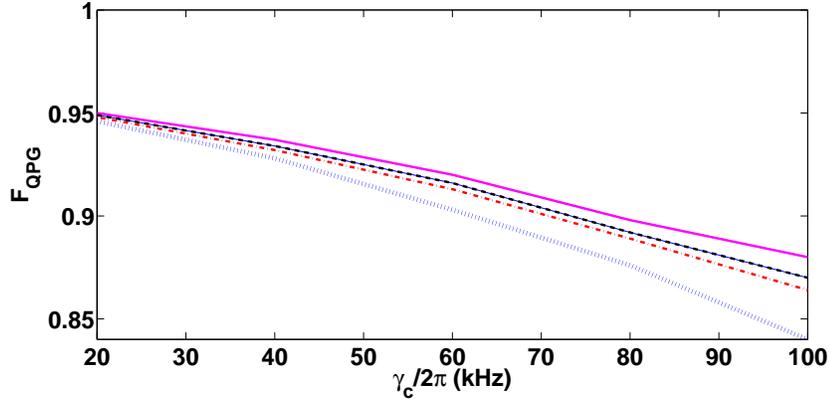}
\caption{  Average fidelity for QPG gate $(F_{QPG})$ calculated from Eqn. \ref{FidelityQPG} for $\phi=\pi/4$ (continuous), $\pi/3$ (dashed), $\pi/2$ (dot-dashed) and $\pi$ (dotted).  We use $\omega_m/2\pi=15.9 \times 10^6$ Hz, $g/2\pi=242\times 10^3$ Hz,$\gamma_m/2\pi=150$ Hz, $\chi=\omega_m, \frac{\omega_c}{g}=\frac{2(p+q)\pi+\phi}{2(q-p)\pi+\phi}$ with $p=2499,q=2500$. }
\label{AverageFidelityQPG}
\end{figure}

Fig. \ref{AverageFidelityQPG} shows the average fidelity calculated from Eqn. \ref{FidelityQPG} for $\phi=\pi/4$ (continuous), $\pi/3$ (dashed), $\pi/2$ (dot-dashed) and $\pi$ (dotted). The average is taken over the ensemble of initial  states given in Eqn. \ref{inputQPG}, where $c_i$'s are randomly chosen.  We set $\omega_m/2\pi=15.9$ MHz, $g/2\pi=241$ kHz, $\gamma_m=150$ Hz and $\omega_c=g\frac{2(p+q)\pi+\phi}{2(q-p)\pi+\phi}$ with $p=2499,q=2500$. It can be seen from the figure that the fidelity depends on phase $\phi$ in the presence of dissipation. This comes from the fact that the ratio $\omega_c/g$ and thereby the gate operation time, depends on $\phi$. If $\phi$ is large, the gate operation time is also large, and as a result system suffers large dissipation. This reduces the fidelity.\\


We calculate $F_{CNOT}$ and $F_{QPG}$ by using the recently achieved experimental parameters and tabulated them in Table.1. The cavities that are used in thesse experiments have Q-factors of the order of $Q\sim 10^3-10^4$ and hence the fidelities for quantum gates are small. However, optomechanical cavities have already been fabricated with a large $Q$-factor $\sim 10^6$ \cite{Sun, Thompson, Chang2}. Using this value we can obtain $F_{CNOT}$ and $F_{QPG}$ respectively to be $\sim 0.975$ and $0.982$ by setting $\omega_m/2\pi=15.9$ MHz, $g/2\pi=241$ kHz, $\omega_c/2\pi\sim$2 GHz and $\gamma_m/2\pi=150$ Hz.\\ 
\begin{table}[h!]
\begin{center}
\label{ExpTable}
\begin{tabular}{|c|c|c|c|c|c|c|c|}
\hline
 Reference & $\omega_m/2\pi$ & $\gamma_m/2\pi$  & $\gamma_c/2\pi$ & $g/2\pi$ & $F_{CNOT}$ & $F_{QPG} (\phi=\pi/8)$ \\
 \hline
\hline
Lecocq \textit{et.al} \cite{Lecocq} & 15.9 MHz & 0.15 kHz  & 163 kHz & 242 kHz & 0.88 & 0.81  \\
\hline
Groeblacher \textit{et.al} \cite{Groeblacher}  & 1 MHz & 0.14 kHz  & 215  kHz & 325 kHz & 0.85 & 0.79 \\
\hline
Murch \textit{et.al} \cite{Murch} & 0.04 MHz & 1 kHz  & 660  kHz & 600 kHz & 0.7 & 0.73 \\
 \hline
\end{tabular}
\caption{Values of the parameters achieved in experiments and calculated quantum gate fidelities. Resonance frequency of the cavity can be calculated using the relation $\omega_c/g=4k-1$ (here $k=2500$) and $\frac{\omega_c}{g}=\frac{2(p+q)\pi+\phi}{2(q-p)\pi+\phi}$ (here $p=2499, q=2500$) for CNOT gate and QPG respectively.}
\end{center}
\end{table}

As mentioned earlier, a photon is required inside the cavity to operate the quantum gate. If the photon leaks out from the cavity before the time $T_G$ then the target state can not be achieved and the fidelity will become small.  Let the mean life-time of photon inside the cavity is $\tau\sim 1/\gamma_c$. At this time, the probability of detecting the photon inside the cavity is $\sim 1/e$. If $\tau> T_G$ then there will be very high probability to reach the target state.  Hence, the condition $T_G \approx \tau$ \textit{i.e.,} the photon life-time is nearly equal to the gate operation time, may provide a gate fidelity above which the quantum gate works properly. For an optomechanical cavity with $\omega_c/2\pi \sim $ GHz, $g/2\pi=242 $ kHz, $\omega_m/2\pi=15.9$ MHz, $\gamma_m/2\pi=150$ Hz, these fidelities are $F_{CNOT}=0.7195$ and $F_{QPG}=0.695$ for $T_G\approx \tau$.


\section{Effect of higher excitation number}\label{HigherStates} 
The eigenstates given in Eqn. \ref{eigenstate}$(a)-(c)$ are valid in the limit of $\chi/g>>1$. In this limit, contributions from higher excited states of mechanical resonator are negligible. If $g$ is comparable to $\chi$, contributions from the higher excited states are significant. This may bring an additional reduction of the fidelity.\\
\begin{figure}[h]
\centering
\includegraphics[scale=0.3]{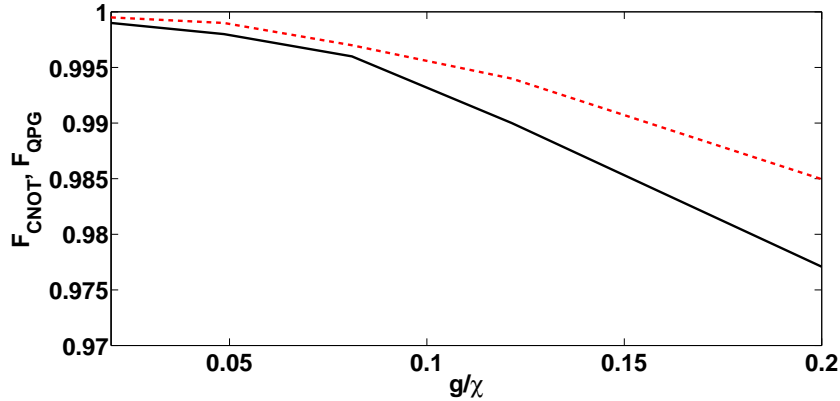}
\caption{ $F_{CNOT}$ and $F_{QPG}$ as a function of $g/\chi$. Continuous line corresponds to $F_{CNOT}$ and dashed line is for $F_{QPG}$ $(\phi=\pi/4)$.  For the CNOT gate, we use $\omega_c/g=(4k-1)$ with $k=2500$. For QPG, we set $\frac{\omega_c}{g}=\frac{2(p+q)\pi+\phi}{2(q-p)\pi+\phi}$ with $p=2499,q=2500$.}
\label{FidelityCNOT&QPGlargeG.eps}
\end{figure}

 In order to see the effect of higher excitation number on the gate fidelities, we plot the average fidelities as a function of $g/\chi$ in Fig. \ref{FidelityCNOT&QPGlargeG.eps}. The continuous line corresponds to the fidelity of the CNOT gate and the dashed line corresponds to the fidelity of the QPG gate $(\phi=\pi/4)$. It is to be noted that the fidelities for both the gate operations decrease as $g/\chi$ increases.  
\section{Summary}\label{Summary}
We proposed a scheme for realizing two-qubit CNOT gate and QPG in an optomechanical system where the mechanical resonator is of Kerr type.  For proper choices of resonance frequencies, Kerr strength and optomechanical coupling strength, the optomechanical cavity behaves like a quantum gate.  The fidelities of the quantum gate operations are very close to unity in the absence of dissipation.  In the presence of dissipation, due to photon loss and mechanical damping, average fidelities of the quantum gate decrease. Using recently achieved experimental values, we show that it is possible to realize CNOT gate and QPG with fidelities of $\sim 0.975$ and $\sim 0.982$ respectively.

Apart from the dissipation, another factor that can reduce the quantum gate fidelities is the coupling strength between the cavity and mechanical resonator. It is seen that the fidelities decrease if the coupling strength is comparable to nonlinear strength. In this limit, perfect phonon blockade does not occur.   
\section{Acknowledgement}
The author acknowledges Indian Institute of Technology Kanpur for postdoctoral fellowship.
\section{References}     
\providecommand{\newblock}{}

\end{document}